\documentstyle[11pt,amsfonts]{article}

\newcommand{\be}{\begin{equation}}
\newcommand{\ee}{\end{equation}}
\newcommand{\bea}{\begin{array}}
\newcommand{\ea}{\end{array}}
\newcommand{\beqa}{\begin{eqnarray}}
\newcommand{\eeqa}{\end{eqnarray}}
\newcommand{\bean}{\begin{eqnarray*}}
\newcommand{\eean}{\end{eqnarray*}}

\def\up#1{\leavevmode \raise.16ex\hbox{#1}}

\def\sqr#1#2{{\vcenter{\vbox{\hrule height.#2pt
        \hbox{\vrule width.#2pt height#1pt \kern#1pt
          \vrule width.#2pt}
        \hrule height.#2pt}}}}

\newcommand{\gapproxeq}{\lower .7ex\hbox{$\;\stackrel{\textstyle
>}{\sim}\;$}}
\newcommand{\lapproxeq}{\lower .7ex\hbox{$\;\stackrel{\textstyle
<}{\sim}\;$}}

\newcounter{appendice}

\def\thebibliography#1{{\bf REFERENCES\markboth
 {REFERENCES}{REFERENCES}}\list
 {[\arabic{enumi}]}{\settowidth\labelwidth{[#1]}\leftmargin\labelwidth
 \advance\leftmargin\labelsep
 \usecounter{enumi}}
 \def\newblock{\hskip .11em plus .33em minus -.07em}
 \sloppy
 \sfcode`\.=1000\relax}

\begin{document}

\centerline{ \LARGE Noncommutative Point Sources }

\vskip 2cm

\centerline{ {\sc    A. Stern}  }

\vskip 1cm
\begin{center}

{\it Department of Physics, University of Alabama,\\
Tuscaloosa, Alabama 35487, USA}

\end{center}

\vskip 2cm

\vspace*{5mm}

\normalsize
\centerline{\bf ABSTRACT} 
We  construct a perturbative solution to classical noncommutative  gauge
theory on 
${\mathbb{R}}^{3}$ minus the origin using the
Groenewald-Moyal star product.  The result describes a noncommutative point charge.   Applying it to the quantum mechanics of the noncommutative hydrogen
atom gives shifts in the $1S$ hyperfine splitting which are first
order in the noncommutativity parameter.    

\vspace*{5mm}

\newpage
\scrollmode
\section{Introduction}

As position eigenstates do not occur in theories with space-space
noncommutativity, there can be no intrinsic notion of a {\it  point} source in
noncommutative physics. It then is argued that point
charges become smeared in noncommutative gauge
theory.\cite{Smailagic:2003rp} 
Gaussian distributions
having   width equal to the noncommutativity scale
were utilized to model  noncommutative sources, and in particular,
sources associated with  noncommutative black holes.\cite{Nicolini:2005vd}

On the other hand,  after going to a star product realization of the
noncommutative algebra, it is possible to show
that point sources persist; i.e., 
there exist nonvanishing solutions to the noncommutative free field equations on 
${\mathbb{R}}^{n}$ minus point(s).
In particular, using  the
Groenewald-Moyal star product, it is easy to construct a perturbative solution to noncommutative $U(1)$ gauge
theory describing a static point charge.  We do this in section
2 for $n= 3$.  Only space-space components of the
noncommutativity tensor affect the fields around the 
 static point source.   A magnetostatic potential is
induced at first order in the  noncommutativity tensor, while
corrections to the electrostatic potential are induced at second
order. These lowest order corrections are  independent of the
choice of star product.  The solution is 
nontrivial in the sense that it is not obtained from a
 Seiberg-Witten map of the commutative Coulumb solution.  The
latter would instead induce a nonvanishing current density away from the
point source.

There is some utility in  applying the lowest order solution to the quantum mechanics of the noncommutative hydrogen
atom.   A  debate in the literature concerns how to treat the
nucleus in the noncommutative theory.\cite{Chaichian:2000si},\cite{Chaichian:2002ew},\cite{Ho:2001aa}  If both the electron and
nucleus are treated on the same footing; i.e., as noncommutative  particles,  their relative
coordinates commute  leading to no   noncommutative corrections to the Coulumb potential.\cite{Ho:2001aa}  On the other
hand, it was argued that the nucleus should be treated as a
commutative object since QCD effects dominate over any  noncommutative
physics.\cite{Chaichian:2002ew}  Corrections then result in the Lamb
shifts due to the noncommutativity of just the electron.  It may be
difficult to answer the debate  conclusively in the absence
of a consistent theory of noncommutative quarks and gluons.  A  pragmatic
approach would be to set bounds on the noncommutativity of the nucleus.
We do this in section 3 by presuming the nucleus to be a noncommutative
point charge in the sense described above.  New shifts result in the hydrogen atom spectra at
lowest order in the noncommutativity parameter,
including in the $1S$ hyperfine  splitting.

Concluding remarks are made in section 4.

\section{Point sources in noncommutative electrodynamics}

   Here we find it helpful to work in terms of SI units, with
$c=1$ (but not $\hbar=1$), where the noncommutative gauge
coupling constant $g_{{}_{SI}}$ has nontrivial units.
Assuming constant noncommutativity $\theta^{\mu\nu}=-\theta^{\nu\mu}$,
the  $U(1)$ gauge
field equations
  read 
\be \partial^\mu F_{\mu\nu} - ig_{{}_{SI}} [ A^\mu, F_{\mu\nu}]_\star = J_\nu\;,\label{ncmxwl}\ee
with
\be F_{\mu\nu}= \partial_\mu A_\nu - \partial_\nu A_\mu
-ig_{{}_{SI}}[A_\mu, A_\nu]_\star \;,\label{fldstrng}\ee
$[\;,\;]_\star$ being the star commutator associated with the
Groenewald-Moyal star
\be \star = \exp\;\biggl\{ \frac {i}2 \theta^{\mu\nu}\overleftarrow{
  \partial_\mu}\;\overrightarrow{ \partial_\nu} \biggr\} \label{gmstr} \ee
We can perturbatively solve these equations starting from the
commutative Coulumb solution in three spatial dimensions
\be A_\mu^{(0)}=- \frac q{4\pi\epsilon_0\; r}\delta _{\mu 0}\;,\label{zordsln}
\ee  $ \mu,\nu,...=0,1,2,3\;$.
Here we included the permittivity constant
$\epsilon_0$ in the current, $ 
J^{(0)}_\mu=(q/\epsilon_0)\delta _{\mu 0}\delta^3(x)$.
  From (\ref{ncmxwl})
 it then follows that
\be g= \frac{g_{{}_{SI}} q}{4\pi\epsilon_0}\label{dmnslsg}\ee is a dimensionless factor. 
Take (\ref{zordsln}) to be the zeroth order term in a Taylor expansion in
$\theta^{\mu\nu}$
\beqa  A_\mu&=&A_\mu^{(0)}+A_\mu^{(1)}+A_\mu^{(2)}+\cdot\cdot\cdot
\eeqa
Assume that the noncommutative current $ J_\mu$ vanishes everywhere away from the origin at all
orders in  $\theta^{\mu\nu}$
\be  J_\mu = 0\;,\qquad r\ne 0 \ee
   Then
(\ref{ncmxwl}) gives
\beqa  \nabla^2 A_0^{(1)}&=& 0\cr & &\cr ( \nabla^2-\partial_0^2)
A_i^{(1)} + \partial_0\partial_i  A_0^{(1)} -\frac{g 
q}{4\pi\epsilon_0}\; \frac{\theta^{ij} x_j}{r^6}&=&0 \;,\qquad r\ne 0\;, \label{fostcfe}
 \eeqa  $ i,j,...=1,2,3\;,$  after extracting the
 first  order terms and applying the
Coulumb gauge $ \nabla\cdot \vec A=0\;$.   A static first order solution is
\be   A_i^{(1)}=  \frac
{g q}{16\pi\epsilon_0} \;\frac{\theta^{ij} x_j}{r^4}\;,\label{fostcsln}
\ee with $ A_0^{(1)}=0 $.  (\ref{fostcsln}) satisfies
the Coulumb gauge condition due to the antisymmetry of $\theta^{ij}$,
and implies the existence of a noncommutative magnetic field
\be B_i^{(1)}=\frac12 \epsilon_{ijk}F^{(1)}_{jk}= -  \frac
{g q}{16\pi\epsilon_0}\;\epsilon_{ijk}\biggl\{\frac{\theta^{jk}}{r^4}- 4
\frac{\theta^{j\ell}x_\ell x_k}{r^6}\biggr\}\ee
We call (\ref{fostcsln}) the inhomogeneous solution.  It
falls off faster than a magnetic dipole potential  \be 
\frac{ \epsilon_{ijk} m_j^{(1)} x_k}{r^3} \;, \label{mdptnl}
\ee yet it cannot be expressed in terms of a  magnetic quadrupole potential
 \be \frac{{\cal M}_{ijk}^{(1)} x_jx_x}{2r^5}\label{qdptnl}\;, \ee
 with constant coefficients  ${\cal M}_{ijk}^{(1)}$.
On the other hand, (\ref{mdptnl}) and (\ref{qdptnl}), along with
higher moment potentials, can be regarded as  homogeneous
terms which can be added to (\ref{fostcsln}).  The moments are
arbitrary, except for being linear in
$\theta^{\mu\nu}$. (For instance, one can have $m_i^{(1)}\propto
\theta^{i0}$ or $\epsilon_{ijk}\theta^{jk}$.)   Additional homogeneous
 terms can be  introduced  with a multi-moment expansion for the time
 component of  $A_\mu^{(1)}$:
\be A_0^{(1)}=-\frac1{4\pi\epsilon_0}\biggl\{ \frac { q^{(1)}} r+ \frac{p_i^{(1)} x_i}{r^3}+ \frac{Q_{ij}^{(1)} x_ix_j}{2r^5}+\cdot\cdot\cdot\biggr\}\;,\label{homoA0} \ee
where the constant  coefficients $ q^{(1)}$, $p^{(1)}_i$, $Q_{ij}^{(1)},...$ are
 undetermined, except that they are linear in
$\theta^{\mu\nu}$.

The first order solution   (\ref{fostcsln}) can be re-expressed in
terms of the zeroth order solution (\ref{zordsln}) and its
derivatives:
\be  A_i^{(1)}= -\frac 14 g_{{}_{SI}}\theta^{ij}  {
   A}^{(0)}_0 \partial_j  {
   A}^{(0)}_0 \ee
This is not a Seiberg-Witten map\cite{Seiberg:1999vs} of $ {
   A}^{(0)}_0$, as commutative gauge transformations ${
   A}^{(0)}_0\rightarrow {
   A}^{(0)}_0+\partial_0\lambda$ do not induce noncommutative gauge
 transformations in $A_\mu$.  The standard expression for the Seiberg-Witten map at first
 order\footnote{Homogeneous terms  ${\cal H}_{{\cal A}^{(0)}_\mu} $,  satisfying
$ {\cal H}_{{\cal A}^{(0)}_\mu+\partial_\mu\lambda}
-{\cal H}_{{\cal A}^{(0)}_\mu} \;=\;\theta^{\rho\sigma}\partial_\rho \lambda\partial_\sigma{\cal H}_{{\cal
    A}^{(0)}_\mu} $ at first order, can be added to this expression.\cite{Asakawa:1999cu},\cite{Jurco:2001rq},\cite{Pinzul:2004tq}} 
\beqa  A_\mu^{(0)}\rightarrow  A^{SW}_\mu &=&{ A}^{(0)}_\mu \;-\;\frac
12g_{{}_{SI}}\theta^{\rho \sigma}\;{ A}^{(0)}_\rho
 \;(\partial_\mu { A}^{(0)}_\sigma
 -2\partial_\sigma A_\mu^{(0)})+\cdot\cdot\cdot\;\label{foswmp}\eeqa
instead leads to a nonvanishing first order  current density
 away from the origin (in addition to a  singular current density at the origin).  Substituting  (\ref{fostcsln}) in
(\ref{foswmp}) gives \be A_\mu^{SW}=- \frac q{4\pi\epsilon_0\; r}\Bigl( 1 +g \frac
{\theta^{0i}x_i}{ r^3} +\cdot\cdot\cdot\Bigr) \delta _{\mu
  0}\;, \ee which is associated with a nonvanishing current density
for $r\ne 0$
\be J_0^{SW}=-\frac {qg}{\pi\epsilon_0}\;\frac
{\theta^{0i}x_i}{ r^6} +\cdot\cdot\cdot \qquad  J_i^{SW}=-\frac {qg}{4\pi\epsilon_0}\;\frac
{\theta^{ij}x_j}{ r^6} +\cdot\cdot\cdot\;,\qquad r\ne0\; \ee 

It is straightforward to extend the inhomogeneous solution to
higher  orders.
At second order  in
$\theta^{\mu\nu}$ the
field equation (\ref{ncmxwl}) gives
\beqa  \nabla^2 A_0^{(2)}+\frac{qg^2}{8\pi\epsilon_0}\;\biggl\{
\frac{{\rm Tr}\theta^2}{r^7}- 7 \frac{[\theta^2]^{ij} x_ix_j}{r^9} \biggr\} &=& 0\cr & &\cr ( \nabla^2-\partial_0^2)
A_i^{(2)} + \partial_0\partial_i  A_0^{(2)}&=&0 \;,\qquad r\ne 0\;, \label{fostcfe}
 \eeqa
in the Coulumb gauge.  It is solved by \be  A_0^{(2)}=-\frac{qg^2}{16\pi\epsilon_0}\;\biggl\{\frac{{\rm Tr}\theta^2}{5r^5}- \frac{[\theta^2]^{ij}
  x_ix_j}{r^7} \biggr\}\label{socep}
\;,\ee and $A_i^{(2)}=0$.  $A_0^{(2)}$ then falls off faster than an electric
quadrupole potential, but cannot be expressed as an octopole
potential.

  The lowest order corrections to the Coulumb potential,
(\ref{fostcsln}) and   (\ref{socep}), were computed using the leading
order of the star commutator.  They are therefore independent of the
choice of star product.

\section{Another look at the noncommutative hydrogen atom}

Now consider a `noncommutative' electron moving in the potential found in the previous section.
Following \cite{Chaichian:2000si} its quantum algebra is defined by
\beqa [\hat x_i,\hat x_j]&=&i\theta_{ij} \cr
[\hat x_i,\hat p_j]&=&i\hbar \delta_{ij} \cr
[\hat p_i,\hat p_j]&=&0\;, \eeqa
along with the usual spin algebra.  It is well known that this can be mapped to the standard Heisenberg algebra, spanned by
$\hat X_i$ and $\hat P_j$,  using
\be  \hat x_i \rightarrow \hat X_i=  \hat x_i + \frac 1{2\hbar} \theta_{ij}\hat
p_j \qquad \hat p_i \rightarrow \hat P_i= \hat p_i\label{cancx}\ee
For the dynamics  in a noncommutative gauge field  we can adapt the standard
Hamiltonian for a   nonrelativistic electron
\be \hat H= \frac 1{2m} \Bigl(\hat p_i- q A_i(\hat x)\Bigr)^2 +
qA_0(\hat x) - \frac{2\mu_B} \hbar \vec{  S}\cdot \vec B(\hat x)\;,\label{nrlham} \ee
where $\mu_B = q\hbar /{2m}$.
Alternatively, $\hat H$ can be realized in terms of differential
operators acting on wavefunctions on 
${\mathbb{R}}^{3}$, using the   
Groenewald-Moyal star (\ref{gmstr}).  For example, the first term 
corresponds to
\be -\frac{\hbar^2}{2m} D_{\star i} D_{\star i} \;\ee
The covariant derivative $D_{\star i}$ must be the same as that
entering in the field equations  (\ref{ncmxwl}) and the definition field strength (\ref{fldstrng}),
here written in the fundamental representation, i e.,
\be  D_{\star i}=\partial_i - ig_{{}_{SI}}  A_i\star  \ee
In comparing with (\ref{nrlham}) one gets the identification of
$g_{{}_{SI}}$ with $q/\hbar$, or equivalently, the
dimensionless coupling constant $g$ defined in  (\ref{dmnslsg}) with the fine
structure constant:
\be g = \frac{q^2}{4\pi\epsilon_0 \hbar} = \alpha \ee

Next we substitute the solution  for $ A_\mu$ and $\vec B$ found in the previous section, keeping only the first order
correction.  The result is 
\be \hat H= \frac 1{2m} \biggl(\hat p_i- \frac
{\alpha^2\hbar }{4} \;\frac{\theta^{ij}\hat x_j}{\hat r^4} \biggr)^2-
\frac {\alpha \hbar}{\hat r}  +    \frac
{\alpha^2\hbar }{4 m} \;\epsilon_{ijk} 
S_i\biggl\{\frac{\theta^{jk}}{\hat r^4}- 4
\frac{\theta^{j\ell}\hat x_\ell \hat x_k}{\hat r^6}\biggr\}   \;, \ee
where $\hat r^2 = \hat x_i \hat x_i$.    Since we  only are interested
in the first order in $\theta$, it doesn't matter if we express the vector
potential and magnetic field as  functions  of  the commuting or noncommuting
coordinates, $\hat X_i$ or $\hat x_i$.  This of course is not the case
for the Coulumb potential. Following \cite{Chaichian:2000si}, it
 can be re-expressed in terms of $\hat X_i$ using (\ref{cancx}).  Thus 
\beqa \hat H&=&\hat H^{(0)} + \hat H_1^{(1)} + \hat H_2^{(1)} + \hat
H_3^{(1)}+\cdot\cdot\cdot\;,\cr & &\cr& & \hat H^{(0)}= \frac 1{2m} \hat P_i\hat
P_i-
\;\frac {\alpha\hbar}{\hat R} \cr & &\cr & &  \hat H_1^{(1)}=\;-\;\frac {\alpha }{2 }\;\frac { \vec \theta \cdot
  \vec L}{ \hat R^3 }\qquad\qquad \hat H_2^{(1)}= \frac{\alpha^2\hbar}{4m}\; \frac { \vec \theta \cdot \vec L}{\hat R^4}\cr & &\cr & &
H_3^{(1)}=\frac{\alpha^2\hbar }{2m}\biggl\{\frac {2(\vec X\cdot \vec S)(\vec X\cdot \vec
  \theta)}{\hat R^6} - \frac { \vec \theta \cdot
  \vec S}{ \hat R^4 }\biggr\}  \;, \eeqa 
where $\hat R^2 = \hat X_i \hat X_i$, $\theta_{ij} =
\epsilon_{ijk}\theta_k$ and the dots indicate higher orders.  $\hat H_1^{(1)}$ was obtained in
\cite{Chaichian:2000si}, while  $\hat H_2^{(1)}$ and  $\hat H_3^{(1)}$
are the new  corrections following from $  A_i^{(1)}$, and are due to
the noncommutativity of the source.\footnote{ It was argued in
  \cite{Ho:2001aa} that the relative coordinate $\hat x_i$ is
commuting, and   that as a result the correction   $\hat H_1^{(1)}$ to
the Coulumb interaction is absent.   On the other
hand, the perturbations  $\hat H_2^{(1)}$ and  $\hat H_3^{(1)}$
persist when   $\hat x_i$ is
commuting, resulting in first order shifts in the hydrogen atom spectrum.}  The latter contains couplings of the noncommutativity to
both the orbital and spin angular momentum,
respectively.
 Corrections to the Lamb shifts 
of the $\ell\ne 0$ states result from   $\hat H_1^{(1)}$ and  $\hat
H_2^{(1)}$.  The matrix
elements are diagonalized by  taking $\vec \theta
=(0,0,\theta)$.  The former
  were computed in \cite{Chaichian:2000si}.    Similar expressions
  result  for the latter.  For the two $2P_{1/2}$ states:
\beqa <\hat H_1^{(1)}>_{2P_{1/2}^{\pm 1/2}}&=&-\;\frac {\alpha\theta
}{2 }\;\biggl<\frac {  L_z}{ \hat R^3 }\biggr>_{2P_{1/2}^{\pm
    1/2}}\;=\;\mp\;\frac {\alpha\hbar\theta }{72 a_0^3
}\;\;\;\;=\;\;\mp\;\frac{\alpha^4m\theta}{72\lambda_e^2}\label{spltone} \\ & &\cr
 <\hat H_2^{(1)}>_{2P_{1/2}^{\pm 1/2}}&=&\;\;\;\;\frac{\alpha^2\hbar\theta }{4m}\;\biggl<\frac {  L_z}{ \hat R^4 }\biggr>_{2P_{1/2}^{\pm 1/2}}=\;\pm\frac{\alpha^2
\hbar^2\theta}{144ma_0^4}=\;\pm\;\frac{\alpha^6m\theta}{144\lambda_e^2}\;\label{splttwo}
\;,\eeqa
  using spectroscopic notation $n\ell_{j}^{m_j}$.  The new
  contribution (\ref{splttwo}) is down by a factor of $\alpha^2$ and thus gives a much
  weaker bound on $\theta$.  According to
  \cite{Eides:2000xc} the current theoretical accuracy on the $2P$ Lamb
  shift is about $0.08$ kHz.   From the splitting (\ref{spltone}), this then gives the following bound on $\theta$
 \footnote{There was a computational error  in the original version of \cite{Chaichian:2000si}.}
\be \theta\; {}^<_\sim \; (6\; {\rm GeV})^{-2}\label{bndntht}\;,\ee
while from
 (\ref{splttwo}) one gets 
\be  \theta\; {}^<_\sim\;(30\; {\rm  MeV})^{-2} \label{srcbndntht}\ee
As has been argued in \cite{Ho:2001aa},  noncommutativity is not the
same for all particles in noncommutative quantum mechanics.
Here (\ref{bndntht}) is a bound on the noncommutativity associated with the test
charge (electron), while (\ref{srcbndntht}) is effectively a bound on  the
lowest order noncommutativity of the source (proton).
Comparing (\ref{srcbndntht}) with
the QCD scale  $\Lambda_{QCD}\sim 200$ MeV,  one
cannot here conclude that  strong interactions
dominate over any noncommutative effects of the source.

More interesting are the matrix elements of
 $\hat H_3^{(1)}$, as they induce new splittings in the $1S$ states,
 thus affecting the hyperfine structure.\footnote{The $\ell=0$ matrix
 elements would vanish with the  addition of a term to  $\hat H_3^{(1)}$ whereby the factor $2$ in braces is changed to $3$.  The origin of such
 a term however is unclear.}\footnote{Noncommutative corrections to the
 $1S$ hyperfine splitting were
 examined previously in \cite{Alavi:2006ss} by expressing the
 dipole-dipole interaction in terms of the noncommutative coordinates
 $\hat x_i$.  Those corrections, however,  go like $\theta^2$ at  the
 lowest order.}  Actually, with the restriction to static  point sources,
 the $1S$ matrix elements are linearly divergent!  To get a finite answer we take into account the finite size of the
 nucleus and  insert the $\Lambda_{QCD}$
 cutoff\footnote{This of course would not be valid for the muonium
 atom ($e^-\mu^+$). Relaxing  the assumption of static sources,
 thereby taking into account recoil effects, may
 cure the ultraviolet divergence
  for that case.}
\beqa <\hat H_3^{(1)}>_{1S_{1/2}^{\pm 1/2}}&=&\frac{\alpha^2\hbar }{2m}\;\biggl<\frac {S_i\theta_j}{ \hat R^6 }\;(2\hat X_i \hat X_j
 -\hat R^2 \delta_{ij})
\biggr>_{1S_{1/2}^{\pm 1/2}}\cr & &\cr & &\cr&=& -\;\frac{\alpha^2\hbar \theta}{6m}\;\biggl<\frac{S_z}{ \hat R^4 }\biggr>_{1S_{1/2}^{\pm
 1/2}}\;=\;\mp \frac{\alpha^2\hbar\theta}{3ma_0^3\Lambda^{-1}_{QCD}}\;=\;\mp\frac{\alpha^5m\theta}{3\hbar\lambda_e\Lambda^{-1}_{QCD}}\;,\label{ncrtnsthpf}
\eeqa
where again $\vec \theta
=(0,0,\theta)$ and we used $\bigl< {\hat X_i \hat X_j}/{ \hat R^n }\bigr>_{\ell=0} =\frac 13\delta_{ij}\;\bigl< {1}/{ \hat R^{n-2}
 }\bigr>_{\ell=0}\;$.  These terms should then mix with the usual $1S$
 hyperfine matrix elements.   According to
  \cite{Eides:2000xc} the current theoretical accuracy on the $1S$ 
  shift is about $14$ kHz.  From the splitting (\ref{ncrtnsthpf}),
 this gives
\be \theta\; {}^<_\sim \; (4\; {\rm GeV})^{-2}\;,\ee for the  noncommutativity of the proton,
which is now well above the QCD scale.  However, without have a
  treatment of noncommutative QCD, the insertion of the QCD
 cutoff in this approach remains uncertain.

\section{Concluding Remarks}

We have found that the
noncommutativity of the electron and proton have distinct
experimental signatures in the hydrogen spectrum.  We further found the same order of magnitude for their bounds. 

There appear to be a number of possibilities for  generalizations of
this work:  a)  One is to obtain the exact solution for the
noncommutative potential and also its dependence on the choice of star
product.  b) Another is to
drop the restriction of static sources.  This will allow for the study
of recoil effects in noncommutative quantum systems.  As stated
earlier, this appears necessary to remove the divergence in the
correction to the $1S$
state of the noncommutative muonium atom.   c) A self-consistent
dynamics for these point sources, at the classical as well as the
quantum level, is then also of interest.  The classical equations of motion
would be analogous to the Wong equations in Yang-Mills theory.\cite{Wong:1970fu},\cite{Balachandran:1977ub} d) Generalizations to other
gauge  theories, including gravity, should be possible.  For the case of gravity this
should lead to yet another description of noncommutative black holes.\cite{Dolan:2006hv}
e) More challenging perhaps would be an attempt to find analogous
solutions in theories with nonconstant noncommutativity.

\bigskip
\noindent
{\bf Acknowledgment}

\noindent
We thank  Ben Harms, Shahin Jabbari and  Aleksandr Pinzul
for useful discussions.


\bigskip

\bigskip
\begin{thebibliography}{99}

\bibitem{Smailagic:2003rp}
  A.~Smailagic and E.~Spallucci,
  J.\ Phys.\ A  {\bf 36}, L467 , L517 (2003).

\bibitem{Nicolini:2005vd}
  P.~Nicolini, A.~Smailagic and E.~Spallucci,
  Phys.\ Lett.\  B {\bf 632}, 547 (2006).





\bibitem{Chaichian:2000si}
  M.~Chaichian, M.~M.~Sheikh-Jabbari and A.~Tureanu,
  Phys.\ Rev.\ Lett.\  {\bf 86}, 2716 (2001).
 
\bibitem{Chaichian:2002ew}
  M.~Chaichian, M.~M.~Sheikh-Jabbari and A.~Tureanu,
  Eur.\ Phys.\ J.\  C {\bf 36}, 251 (2004).


\bibitem{Ho:2001aa}
  P.~M.~Ho and H.~C.~Kao,
  Phys.\ Rev.\ Lett.\  {\bf 88}, 151602 (2002). 

\bibitem{Seiberg:1999vs}
  N.~Seiberg and E.~Witten,
  JHEP {\bf 9909}, 032 (1999).

\bibitem{Asakawa:1999cu}
  T.~Asakawa and I.~Kishimoto,
  JHEP {\bf 9911}, 024 (1999).

\bibitem{Jurco:2001rq}
  B.~Jurco, L.~Moller, S.~Schraml, P.~Schupp and J.~Wess,
  Eur.\ Phys.\ J.\  C {\bf 21}, 383 (2001).


\bibitem{Pinzul:2004tq}
  A.~Pinzul and A.~Stern,
  Int.\ J.\ Mod.\ Phys.\  A {\bf 20}, 5871 (2005).



\bibitem{Eides:2000xc}
  M.~I.~Eides, H.~Grotch and V.~A.~Shelyuto,
  Phys.\ Rept.\  {\bf 342}, 63 (2001).


\bibitem{Alavi:2006ss}
  S.~A.~Alavi,
  arXiv:hep-th/0608107.

\bibitem{Wong:1970fu}
  S.~K.~Wong,
  Nuovo Cim.\  A {\bf 65S10}, 689 (1970).

\bibitem{Balachandran:1977ub}
  A.~P.~Balachandran, S.~Borchardt and A.~Stern,
  Phys.\ Rev.\  D {\bf 17}, 3247 (1978).

\bibitem{Dolan:2006hv}
For a $3D$ example, see
  B.~P.~Dolan, K.~S.~Gupta and A.~Stern,
  Class.\ Quant.\ Grav.\  {\bf 24}, 1647 (2007).


\end{thebibliography}
\end{document}